\documentstyle[aps,psfig]{revtex}
\begin{document}
\draft
\date{\today}
\title{Towards the Thermodynamics of Localization Processes}
\author{Paolo Grigolini$^{1, 2, 3}$, Marco G. Pala$^{3}$,
Luigi Palatella$^{3 , 4}$ and Roberto Roncaglia$^{4}$}
\address{$^{1}$Istituto di Biofisica del Consiglio Nazionale delle
Ricerche, Via S. Lorenzo 26, 56127 Pisa, Italy}
\address{$^{2}$Center for Nonlinear Science, University of North
Texas, P.O. Box 5368, Denton, Texas 76203-5368}
\address{$^{3}$ Dipartimento di Fisica dell'Universit\`a di Pisa, Piazza
Torricelli 2, 56100 Pisa, Italy}
\address{$^{4}$ Istituto Nazionale per la Fisica della Materia, Dipartimento di Fisica,
via Buonarroti 2, 56126 Pisa, Italy }
\maketitle
\begin{abstract}
We study the entropy time evolution of a quantum mechanical model,
which is frequently
used as a prototype for Anderson's localization. Recently
Latora and Baranger [V. Latora, M. Baranger, Phys. Rev.Lett. {\bf 82},
520(1999)] found that there exist three entropy regimes,
a transient regime of passage from dynamics to
thermodynamics, a linear in time regime of entropy increase, namely
a thermodynamic regime of Kolmogorov kind, and a saturation regime.
We use the non-extensive entropic indicator recently advocated by Tsallis
[ C. Tsallis, J. Stat. Phys. {\bf 52}, 479 (1988)] with a mobile
entropic index $q$, and we find
that with the adoption of the ``magic'' value $q = Q = 1/2 $
the
Kolmogorov regime becomes more extended and more distinct than with
the traditional entropic index $q = 1$.
We adopt a two-site model to explain these properties by means of
an
analytical treatment and we argue that $Q =1/2$
might be a typical signature of the occurrence of Anderson's localization.
\end{abstract}

\section{introduction}

In this paper we focus our attention on the process of localization
discovered by Anderson\cite{anderson1,anderson2}, and we
discuss the corresponding time evolution using
the non-extensive thermodynamics view of Tsallis\cite{tsallis1,tsallis2}.
The subject of Tsallis non-extensive thermodynamics
is attracting the interest of an ever increasing number of
investigators in different branchs of the complexity
theory (see, for instance, Ref\cite{tsallis3}).
We think, however, that the connection
with dynamics, especially quantum dynamics, is not yet
explored with the attention that this subject would require.

To  make easier for the reader to understand the
conceptual difficulty of this problem, it is convenient to make
a short review of this intriguing subject. First of all, we want to remind
the reader that there are two classes of entropic indicators. The
first is that of the indicators expressed in terms of
trajectories, which are consequently confined to the classical case. The
second class is that of the entropic indicators expressed in terms of
either classical or quantum
distributions. The second class is directly connected to the focus
of the present paper, which is in fact devoted to
to a quantum problem. However, we are convinced that there is a subtle
equivalence between the two classes of entropy indicators, and that understanding
some properties of the first class can make easier for us to make a
proper balance on the results obtained in the present paper.

\subsection{Entropies in terms of trajectories}

        The Kolmogorov-Sinai (KS) entropy\cite{kolmogorov,sinai}
is considered to be a dynamic property of a single
classical trajectory measuring the randomness
which is responsible for the thermodynamic properties
of the system under study\cite{zaslavsky}.
In the recent literature there are interesting examples of adoption of
this class of entropic indicators either in the ordinary extensive
form\cite{Gaspard,DAV98} or in the new non-extensive form of
Tsallis\cite{marcoluigi}. We think that
the more recent work of Jin and Grigolini\cite{jin} is of special
interest for the present paper. The stimulus for this work was
given by the heuristic arguments of
Refs.\cite{heuristic1,heuristic2,heuristic3} which provide in fact a
strong motivation for the generalization of the KS entropy. The
authors of Ref.\cite{jin} show that the Tsallis non-extensive
generalization of the KS entropy, referred to as
Kolmogorov-Sinai-Tsallis (KST) entropy, can be expressed
 in terms of an average over the invariant distribution
$p(x)$, thereby implying the stationarity condition.
The explicit form of this KST entropy  is:
 \begin{equation}
  H_{q}(t) = [1 - k(q) \int dx p(x)^{q} \xi(t,x)^{1-q}]/(q-1).
  \label{fundamentalresult}
  \end{equation}
  Note that the departure point of the theory leading to this
  interesting result is given by a repartition of the phase space of
  the system under study into cells of small size. This is
  mirrored by the fact that $k(q) \equiv  (2/l)^{q-1}$, where
  $2/l$ denotes the size of these cells. The function $\xi(t,x)$
  is defined by:
  \begin{equation}
 \xi (t)\equiv \lim_{|{{\Delta} y(0)}| \to
0}\big{|}{\frac{\Delta y(t)}{\Delta y(0)}}\big{|}.
\label{deltadefinition}
\end{equation}
 We are now in the right position to show how the
 detection of the entropic index corresponding to
 the thermodynamic
 nature of the dynamics under study is carried out. Let us assume that
 the sensitivity to the initial condition is expressed by:
 \begin{equation}
 \xi(t) = [1 + (1-Q) \lambda_{Q} t]^{1/(1-Q)}.
 \label{typicalsensitivity}
 \end{equation}
 Let us replace Eq.(\ref{typicalsensitivity}) into
 Eq.(\ref{fundamentalresult}) and let us change the mobile
 index $q$. It is evident that the resulting time evolution of the
 KST entropy is a power law behavior faster or slower than the
 linear in time evolution. A linear in time entropy increase can only be found
 if the entropic index $q=Q$ is adopted.  This is the entropic index
 corresponding to the real nature of the thermodynamics of the system.
 The fact that $Q\neq 1$ means that the thermodynamics of the system
 is not extensive. If it happens that the magic entropic index $Q$ has
 the traditional value $Q=1$, then the connection between thermodynamics
 and dynamics is established in the manner recently pointed out, for
 instance, in the illuminating book by
 Zaslavsky\cite{zaslavsky}. This means a dynamics-thermodynamics
 connection, with dynamics characterized by the ordinary Lyapunov
 coefficients. Note in fact that $Q=1$ means that the sensitivity
 to initial conditions given by Eq.(\ref{typicalsensitivity}) becomes
 the ordinary exponential sensitivity and that the thermodynamic
 nature of the system is
 expressed by 
 the ordinary KS entropy.  It is worth remarking  that at $Q=1$ the KST
 entropy of Eq.(\ref{fundamentalresult}) adheres to the
 prescriptions of the Pesin theorem in its ordinary form\cite{pesin}.

 These interesting properties have been originally pointed out in the
 important work  of
 Refs.\cite{heuristic1,heuristic2,heuristic3}. However, the special form for
 $H_{q}(t)$ of
 Eq. (\ref{fundamentalresult}), in addition to taking
 into account the heuristic arguments of
Refs.\cite{heuristic1,heuristic2,heuristic3}
  serves also the useful purpose of establishing the
 genuine entropic index $Q$ even when dynamics are not exactly
 characterized by the form of Eq.(\ref{typicalsensitivity}), and
 system's dynamics are rather characterized by a distribution of
 power indices \cite{politi,simone}. In this case the search for the proper
 entropic index $Q$ by means of the prescription of
 Eq.(\ref{fundamentalresult}) is especially convenient\cite{note1}.

 The connection between this classical case and the quantum case
 under study in this paper is given by the fact that both cases
 imply $Q < 1$. Before addressing the
 discussion of the consequences of this property
 for quantum diffusion, it is convenient to
 to let the reader know about the meaning of
 $Q \neq 1$ in the case of classical diffusion.
The case $Q >1$ has been discussed
 in Ref.\cite{ANNA} and it has been pointed out
 that it leads to a form of diffusion faster
 than ordinary Brownian motion. Note that in this case
 the sensitivity to initial condition
 is given  by Eq.(\ref{typicalsensitivity}) with $Q>1$. This means
 that two trajectories with very close initial conditions depart one
 from the other so fast as to make their distance diverge at a finite
 time. It is thus plausible that $Q>1$ yields a diffusion faster than
 the ordinary diffusion.

 Much more important for the conclusions of this paper is
 the case where $Q<1$. In the classical case this means a
 localization process occurring within a finite time scale.
As shown in Ref.\cite{simone}, the first step to prove this conjecture
is given by the paper by
 Tsallis and Bukman\cite{buckman}. These authors studied a family of
 non-linear Fokker-Planck equation and proved that the adoption of the
 Tsallis non-extensive entropy yield the exact form of solution, and
 that the case $Q < 1$ corresponds to \emph{subdiffusion}. On the other
 hand, both the KST entropy of Eq.(\ref{fundamentalresult}) and the
 dynamic approach to diffusion\cite{mannella} imply
 the \emph{stationary} assumption. It is shown\cite{simone} that
 \emph{subdiffusion} in a \emph{stationary} condition yields localization
  with a finite time scale. This
theoretical prediction has been fully supported by the numerical results of
 Ref.\cite{simone} on the dynamics of the logistic map at the
 onset of chaos. At the end of this paper we shall come back to
 discussing the consequences that this finding has on quantum
 diffusion, if the assumption is made that a sort of statistical
 equivalence exists between fractal dynamics in classical physics
 and the quantum dynamics of the Anderson Hamiltonian.

\subsection{Entropies in terms of distributions}

 As earlier mentioned, the second class of entropies is expressed  in terms
of the
classical or quantum Liouville
density\cite{zurekpaz,brumer,miller,luigi,baranger},by means of either the
traditional Gibbs-Boltzmann form or the Tsallis \cite{tsallis1}
non-extensive form.
In literature we find the implicit assumption that these
two forms of entropic indicators are equivalent\cite{zurekpaz}.
We take this equivalence for granted but, in principle, we leave
open the choice between the
extensive and non-extensive form. The assumption of equivalence
between the two classes of entropy will afford a criterion to make a
choice. In fact, we note that the KST entropy as well as the ordinary
KS entropy has the meaning of
entropy  increase per unit time: Thus it seems to be
natural to interpret the
regime where the
von Neumann entropy is found to
increase linearly in time as the expression of a genuinely
thermodynamical condition.
This point of view is supported by the interesting results of Zurek
and Paz \cite{zurekpaz}, and by the more recent computer calculations of
other authors \cite{brumer,miller}.

Notice that both Ref.\cite{brumer} and Ref.\cite{miller} support the
view of Zurek and Paz, and that all these
papers\cite{zurekpaz,brumer,miller} establish a
connection between dynamics and thermodynamics in the deep classical
regime. The work of \cite{luigi} has established that the entropy time
evolution studied by Pattanayak and Brumer \cite{brumer} results in a
power law increase if we assign to the Planck constant values
corresponding to the region of transition from the classical to the quantum
regime. The authors of Ref.\cite{brumer} did not consider this
condition to be relevant, probably because it is far from the
condition of linear in time increase, which is correctly judged by them as
the only one with a thermodynamical significance. In Ref.\cite{luigi} also
this
special condition has been recognized as being thermodynamic due to
the adoption of the non-extensive perspective advocated by
Tsallis\cite{tsallis1,tsallis2,tsallis3}:The authors of this
paper\cite{luigi}
pointed out that $q<1$ in this case reflects the
occurrence of Anderson localization.

Another important paper, worth of mention, is that of Latora and
Baranger\cite{baranger}. These authors studied several maps,
characterized by deterministic chaos. This is a remarkable paper
since it corresponds to studying the time evolution of the Gibbs
entropy expressed in terms of the probability distribution, and can
be considered as the classical counterpart of the time evolution of
the von Neumann entropy of Ref.\cite{miller}. These authors find that the
time evolution of the Gibbs entropy is characterized by three time
regimes:(i) an early regime of exponential increase,(ii) an
intermediate-time regime of linear increase, and, finally, (iii), a
saturation regime. The earlier
remarks naturally lead us to conclude, in a full accordance with the
view of these authors\cite{baranger},
that dynamics become compatible with thermodynamics
only in the intermediate-time regime. We shall refer to this
intermediate time regime as Kolmogorov regime, for reasons that are
made clear by the equivalence between the two classes of entropy.

\subsection{Purpose and outline of the paper}

To substantiate the conjecture of Ref.\cite{luigi} on the non-extensive
nature of the thermodynamics of a localization process, we devote
this paper to the numerical treatment
of the Anderson tight-binding Hamiltonian. In other words, we
 study  a
prototype of Anderson localization processes, rather
than the quantum kicked rotator. It is well known that
the saturation of the energy increase of the quantum kicked rotator
is due to quantum correlations, namely, the same cause as
that responsible for Anderson's
localization \cite{felix}. However, the two systems, although
equivalent, are not identical and there might be the doubt
that the remarks of Ref.\cite{luigi} do not apply to the genuine
model of Anderson's localization. Thus, we think that the direct
study of the prototype model of Anderson's localization, if we find
that also in this case the mobile entropic index $q$ must be assigned
a value $Q < 1$, will
make more convincing our conjecture about non-extensive thermodynamics and Anderson's
localization.

The outline of this paper is as follows.
Section II defines the model under study in this paper and
points out that the Anderson noise has the twofold role of creating
statistical mechanics and localization. The former aspect, in an
apparent conflict with the latter, is
compatible with the existence of a thermodynamic perspective.
Section III is devoted to the illustration of the numerical results of
this paper. This numerical treatment
has to be considered as a sort of exact treatment of the entropy
time evolution triggered by Anderson randomness. An analytical
treatment of the problem, shedding light into the reasons why the
Tsallis entropy indicator is a so efficient indicator, is illustrated
in Section IV. Finally, in Section V we aim at making a proper balance
on the results of this paper.

 \section{Master equation, Anderson Localization and Statistical
 mechanics}

   The main purpose of this Section is to show that
   the tight-binding Hamiltonian system that we use to discuss
   Anderson's localization is a remarkable example of joint action of
   randomness and order. This is, in other words, a system which
   is not equivalent to the classical condition of full chaos.
   Rather, as we shall see, this is system equivalent to the
   condition of weak classical chaos, or to the condition of sporadic
   randomness.

   We study a system described by the following Hamiltonian
   \begin{equation}
   H = H_{0} + W ,
   \label{totalhamiltonian}
   \end{equation}
   where
   \begin{equation}
   H_{0} \equiv \sum_{m} E_{m} |m \rangle \langle m|
   \label{unperturbedhamiltonian}
   \end{equation}
   and
   \begin{equation}
    W \equiv V \sum_{m}(|m \rangle \langle m+1|+ |m+1 \rangle \langle m|).
   \label{perturbation}
   \end{equation}

   This is the  Hamiltonian originally taken into account by
   Anderson\cite{anderson1,anderson2}. We make this Hamiltonian
   result in a transport process different from the ballistic
   diffusion of a perfect crystal assuming that
   \begin{equation}
   E_{m} = \epsilon + \phi_{m}.
   \label{energies}
   \end{equation}
   Here we are assuming that with changing site there is a fluctuation
   $\phi_{m}$ around the common value $\epsilon$.We assume no correlation
   among different sites, namely
   \begin{equation}
      \langle\phi_{m}\phi_{m\prime}\rangle = A \delta_{mm\prime}.
   \label{nocorrelation}
   \end{equation}

   It has to be pointed out that according to the prescriptions of
   quantum statistical mechanics, any entropy indicator must be
   expressed in terms of the density matrix associated to the
   Hamiltonian of Eq.(\ref{totalhamiltonian}). The time evolution
   of this density matrix
   is unitary, and consequently any form of entropic indicator,
   expressed in terms of the density matrix, is time independent.
   From this point of view, there is no difference between the
  system under study and a system characterized by regular dynamic
   properties, and thus strongly departing from the randomness
   condition intuitively associated to the second principle.

   This is disconcerting. To a first sight, in fact,
    the Anderson prescriptions of Eqs.(\ref{energies}) and
   (\ref{nocorrelation}) sow seed of randomness into the system
   dynamics and the entropy indicator should make this randomness
   ostensible. So the question is raised of how to make the entropy
   indicator sensitive to this randomness.

 The source of entropy increase in the model of  Zurek and Paz\cite{zurekpaz}
 is given by the deterministic chaos that the
 system would exhibit in the classical limit. To
 trigger a regime of entropy increase, these authors took into
   account the influence of the environment
    as a source of dephasing, a process that does not imply any
    exchange of energy between the system and its environment. This is in
    line with the second principle which forces entropy to
    increase (or to remain constant if the process is reversible) if no
    thermal exchange with the environment is allowed. This means that
    an interaction between system and environment is
    allowed, provided that it does not cause any energy exchange. The
    perspective of Zurek and Paz is not trivial: This is so because,
    even if the entropy increase is made possible by the key
    ingredient of external fluctuations, these are so weak that
    the time scale of the process of transition from dynamics to
    thermodynamics is determined by the Lyapunov coefficients,
    namely, a genuinely dynamic property of the Hamiltonian system
    under study.

    Here we find that the Anderson randomness is a sort of
    counterpart of the deterministic chaos randomness of
    Zurek and Paz\cite{zurekpaz}.  To reveal this randomness we must
    adopt the statistical density matrix defined by:
    \begin{equation}
    \rho_{S}(t) \equiv \int d[\phi] w([\phi]) \rho([\phi],t).
    \label{statdensity}
    \end{equation}
    Any contribution to the integral of Eq.(\ref{statdensity}),
    $\rho([\phi],t)$, is an ordinary density matrix corresponding to a
    given random distribution of the energy fluctuations
    $\phi_{m}$. The symbol $[\phi]$ denotes a given Anderson
    realization, namely, $[\phi] \equiv \phi_{1},
    \phi_{2},\ldots\phi_{i}\ldots$.
    Note that, as a consequence of the assumption of
    Eq.(\ref{nocorrelation}), we have
    \begin{equation}
    w([\phi]) =
    \ldots p(\phi_{m-1})p(\phi_{m})p(\phi_{m+1})\ldots
    \label{factorization}
    \end{equation}
   We make the assumption that the random distribution of the site
    energies follows the Cauchy prescription

   \begin{equation}
   p(\phi) = \frac {1}{\pi} \frac {\gamma}{\gamma^{2} + \phi^{2}},
   \label{cauchy}
   \end{equation}
   where $p(\phi)$ denotes the probability that the energy of a given
   site fluctuates by the quantity $\phi$ about the common value
   $\epsilon$ .
    It is interesting to remark that the time evolution of the
    average density matrix $\rho_{S}(t)$ of Eq.(\ref{statdensity})
    becomes identical to that of a perfect lattice if $\gamma = 0$.
   If, on the contrary, the value of the parameter
   $\gamma$ increases, the time evolution of the
   statistical density matrix $\rho_{S}(t)$ increasingly departs from
   the prescription of  unitary time evolution.
   Consequently, $\gamma$, the width of the Anderson noise,
   can be regarded as the randomness intensity
   of the system.

   We are therefore in a position to establish a comparison
   between the time evolution of $\rho_{S}(t)$ and the density matrix
   of the quantum kicked rotor\cite{miller,luigi}. In
   the latter case the source of randomness is given by the
   deterministic chaos of the classical time evolution of the system.
   In both cases, randomness has a twofold role. In the early time
   region of
   the process this randomness creates a condition of transport
   similar to that of ordinary Brownian motion.At later times, the
   diffusion
   process is quenched by the
   occurrence of the Anderson localization.

   The model under study rests only on the two parameters
  $\gamma$ and $V$, and it would be tempting for us
  to refer ourselves to the condition:

   \begin{equation}
   \gamma < V.
   \label{weakrandomness}
   \end{equation}
   as $\emph{weak chaos}$. On the same token, we are tempted
   to refer ourselves to the condition:
   \begin{equation}
   \gamma > V.
   \label{strongrandomness}
   \end{equation}
   as a condition of $\emph{strong chaos}$.
   This would be incorrect, since, as we shall see through the joint
   use of numerical calculations (Section III) and analytical theory
   (Section IV), in both conditions the Anderson randomness and the
   quantum correlations are present, and, in a sense, the condition of
   Eq.(\ref{strongrandomness}) has the effect of realizing Anderson's
   localization at earlier times. Thus, both conditions have to be
   considered as being equivalent to the weak chaos of classical
   mechanics. The case of Eq.(\ref{strongrandomness}) makes it possible
   to rest on analytical calculations. Therefore in Section IV
   we shall focus on the
   condition of Eq.(\ref{strongrandomness}).

   The twofold role of the Anderson randomness has been studied, in
   the case $V < \gamma$,  in an
   earlier publication by Mazza and Grigolini\cite{mazza}.
    The authors of this paper
   prove that the time evolution of the site population
   $p_{n}(t)$ is driven by the following generalized master
   equation:
   \begin{equation}
   \frac{\partial}{\partial t}p_{n}(t)
   =-\sum_{m \neq n} \int_{0}^{t}dt^{\prime}
   \Xi_{nm}(t-t^{\prime})[p_{n}(t^{\prime}) - p_{m}(t^{\prime})].
   \label{generalizedmasterequation}
   \end{equation}
   The authors of Ref.\cite{mazza} prove that in the deep regime of
   strong Anderson noise ($\gamma > V$) the memory kernel
$\Xi_{nm}(t-t^{\prime})$
   becomes
   \begin{equation}
   \Xi_{nm}(t) = 2 K(t)(\delta_{n,m^{\prime}+1}
    + \delta_{n,m^{\prime}+1}),
    \label{structure}
    \end{equation}
    where
    \begin{equation}
    K(t) = \frac{V^{2}\gamma}{\gamma^{2} - V^{2}}\frac{1}{\hbar ^{2}}
    [\gamma exp(-2\gamma t/\hbar) - V exp(-2Vt/\hbar)]+\frac{V^{3}}{\pi \gamma \hbar^{2}}\cos(2Vt/\hbar).
    \label{memoryofstrongrandomness}
    \end{equation}
    It is worth stressing that the oscillatory term on the
    $\emph{r.h.s.}$ of Eq.(\ref{memoryofstrongrandomness}) (the second
    term on the   $\emph{r.h.s.}$ of Eq.(\ref{memoryofstrongrandomness}))
    does not play any relevant role and was introduced by the authors
    of Ref.\cite{mazza} for the minor purpose of reproducing the weak
    and fast oscillations revealed by the numerical treatment. In the
    theoretical treatment of Section IV we shall make an
     approximation equivalent to disregarding the influence of this term.

    The main result of Ref.\cite{mazza} is that in the time region
    \begin{equation}
    \frac{\hbar}{\gamma} << t << \frac{\hbar}{V}
    \label{brownianregion}
    \end{equation}
    the time evolution of the system is virtually indistinguishable
    from ordinary Brownian diffusion. This is so because the negative
    and slow exponential appearing in
    the $\emph{r.h.s.}$ of Eq.(\ref{memoryofstrongrandomness}))
    is not yet strong enough as to balance the fast and strong
    exponential.
    In the case $\gamma >> V$ Anderson's localization occurs at the
    time $\hbar/V$.
    Note that the condition of ordinary statistical mechanics is
    recovered for $V\rightarrow 0$, while keeping the quantity
    $V^{2}/\gamma$ constant. In this limiting condition the
    slow negative tail of Eq.(\ref{memoryofstrongrandomness})
     can be neglected, and the
    resulting statistical process is indistinguishable from
    that where the dephasing process is due to the environment. In
    this case, as we shall see in Section V, we recover the condition
    of ordinary statistical mechanics, denoted by $Q = 1$.

   \section{numerical results}
  The numerical calculations have been done producing, first of all,
  about $1,000$ realizations of the Hamiltonian system of
  Eqs.(\ref{totalhamiltonian}) ,(\ref{unperturbedhamiltonian}) and
  (\ref{perturbation}). We have also checked that in the time span
  illustrated by Figures in this letter the result is not changed if
  the number of realizations is increased.
  Each realization is obtained by using a
  random noise generator which assignes to any site $|m\rangle$ a fluctuation
  $\phi_{m}$ so as to realize the Cauchy prescription of
  Eq.(\ref{cauchy}). The Hamiltonian of each realization is
  diagonalized so as determine the corresponding time evolution and
  the corresponding density matrix.
  For any realization the initial condition is given by the
  wave function $|\psi(0)\rangle = |m=0\rangle$. Finally an average over all the
  realizations is made.

\begin{figure}
\centerline{
\vbox{
\psfig{figure=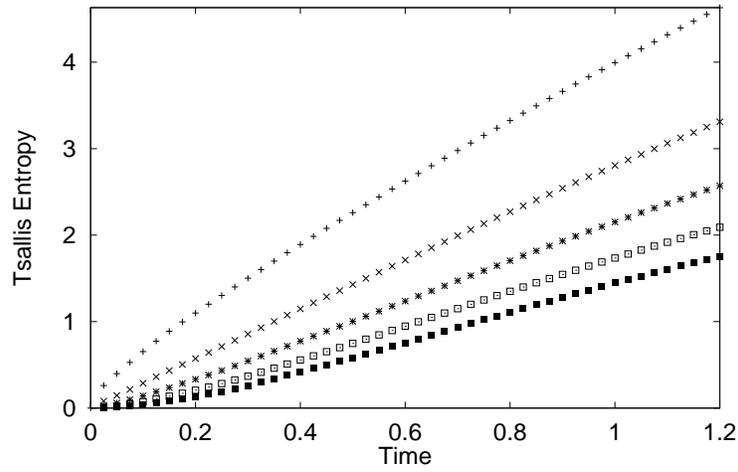,width= 3.8 in}
}}
\caption[]{
 
  The entropic indicators $S_{q}(t)$ as a function of time.
  The values of the parameters are: $\gamma/\hbar= 0.3$ and $V/\hbar = 1$.
  The plotted curves from the bottom to the top refer to: $q = 0.3, q =
  0.4, q =0.5, q =0.6$ and $q = 0.7$.
}
\label{quattro}
\end{figure}

  The numerical results concern these two distinct cases: (i)
  $\gamma>>V$ and (ii) $\gamma<<V$. Fig.\ref{uno} refers to case (i) and Figs.
  \ref{due},\ref{tre} and \ref{quattro} to case (ii). Let us examine case (i) first.
  Fig.\ref{uno} shows the second moment of the distribution
  $M_{2}= \sum_{m} L^{2}p_{m}(t)$. $L$ denotes the lattice
  spacing and for simplicity we assume $L =1$.
  We see that the time evolution of $M_{2}(t)$ undergoes a non linear
  time increase for a time interval of the order of $\hbar/\gamma$.
  After this first time region, the increase of $M_{2}(t)$ becomes
  linear in time. The regime of linear increase lasts for a time of
  the order of $\hbar / V$. In the last time regime the function
  $M_{2}(t)$ tends to become time independent, thereby signalling the
  occurrence of Anderson's localization.

  Fig.\ref{uno} illustrates the time evolution of the two entropy
  indicators, $S_{1}(t)$ and $S_{1/2}(t)$ as well as that of $M_{2}(t)$.
   In fact, as earlier discussed in detail, we think that
  some interesting information can be derived from the observation
  of the entropy indicator:
  \begin{equation}
  S_{q}(t) \equiv \frac {1}{q-1} (1- Tr \rho_{S}(t)^{q}).
  \label{tsallisindicator}
  \end{equation}
  It is well known\cite{tsallis1} that
  the traditional von Neumann entropy is obtained from
  Eq.(\ref{tsallisindicator}) setting $q = 1$.  Notice that
  the choice of $q = Q =1/2$ is the result of a search that will be
  discussed in Section IV. Here we limit ourselves to noticing that the
  transition to the linear regime with $q = 0.5$ is faster than
  in the case $q = 1$. We also notice that the slope of this entropy
  increase per time unit is $2V/\hbar$, corresponding to the
  theoretical predictions of Section V.In conclusion, the time behavior
  of both $S_{1}(t)$ and $S_{1/2}(t)$ is a reflection of the three time regimes
  revealed by the time evolution of $M_{2}(t)$. With the help of Fig.
  \ref{uno} we see that with $q = 1/2$ the time duration of the Kolmogorov
  regime is slightly more extended than in the case $q = 1$. With the
  help of Fig.\ref{uno} we also see that with $S_{1}(t)$ the linear increase
  is not clearly separated from the regime of
  logarithmic dependence in time, which finally is changed into the
  saturation regime corresponding to the the occurence of Anderson's
  localization.  Thus, $S_{1/2}(t)$ is an indicator of the three
  regimes sligthly more efficient than $S_{1}(t)$. This aspect, as we
  shall see with the help of Fig. \ref{due} and \ref{tre}, is more pronounced
  in case (ii).

  Fig.\ref{due} and \ref{tre} illustrate the same properties as those of
  Fig.\ref{uno}, referred to case (ii).
  We see that the function $M_{2}(t)$ signals a
  transition to the statistical regime of Brownian diffusion at a
  time of the order $\hbar/V$. In the time scale explored by Fig.\ref{due}
  there is no sign of the occurrence of Anderson's localization,
  which takes place at a much later time. Even in this case the
  entropic analysis reveals the existence of the three regimes
  of Latora and Baranger\cite{baranger}. Even in this case the
  Kolmogorov regime lasts for a time of the order
  of $\hbar/V$ and even in this case the
  entropy $S_{1/2}(t)$ is a much more accurate indicator of the
  Kolmogorov regime. Even in this case  the slope of $S_{1/2}(t)$
  in the regime of linear increase is
  given by $2V/\hbar$. We note that at a time of the order
  of $\hbar/V$ the entropy $S_{1/2}(t)$ makes a transition
  to a regime of logarithmic dependence on time. This aspect is made clear by
  Fig.\ref{tre}, whose abscissas are expressed in a logarthmic scale for that 
  purpose.
  Again the adoption of $S_{1}(t)$ as entropy indicator does not establish a clear
  distinction betwen the regime of linear increase and that of
  logarithmic dependence on time.

  Of some interest is also Fig.\ref{quattro}, which shows the time evolution of
  $S_{q}(t)$ for several values of the mobile entropic index $q$. We
  see that all these entropic indicators signal a transition to the
  regime of logaritmic dependence on time at times of the order of
  $\hbar/V$. We also note in the first and second time regime a pattern
  of curves reminding the form of a leaf. This leaf effect will be
  discussed again in Section IV. We also note that this leaf effect is
  reminiscent of that revealed by the study of the kicked quantum
  rotor of Ref.\cite{luigi}. The authors of Ref.\cite{luigi} made the
  conjecture that this leaf effect might be related to the occurrence
  of Anderson's localization. With the help of the numerical results
  of this Section and of the theoretical analysis of Section IV, in the
  concluding remarks of Section V we shall address again the
  discussion of this intersting issue.

   \section{the non-extensive interpretation: an analytical treatment}

   The numerical analysis of Ref.\cite{mazza}, suppported also by the
   numerical treatment of this paper, proves that the deep regime of
   strong Anderson randomness is characterized by the important fact
   that the transition from $m$ to $m+1$ is statistically independent
   of that from $m$ to $m-1$. This means that it is possible to carry
   out an analytical treatment based on the study of only two sites.
   The adoption of the distribution of Eq.(\ref{cauchy}) yields the
   following values for the four elements of the
   statistical density matrix of Eq.(\ref{statdensity})
   \begin{equation}
   (\rho_{S}(t))_{11}= \int_{-\infty}^{+\infty}dE \frac{1}{\pi}
   \frac{2\gamma}{4\gamma^{2} + E^{2}}[1 -
   \frac{4V^{2}}{4V^{2}+E^{2}}sin^{2}(\sqrt{E^{2} + 4 V^{2}} \frac{t}{2 \hbar})],
   \label{oneone}
   \end{equation}
   and
    \begin{equation}
   (\rho_{S}(t))_{12}= \int_{-\infty}^{+\infty}dE
   \frac{2 i \gamma V}{\pi }\frac{1}{4\gamma^{2} +E^{2}}
   \frac{1}{\sqrt{ E^{2} + 4 V^{2}}} sin(\sqrt{E^{2} + 4 V^{2}} t/\hbar).
   \label{onetwo}
   \end{equation}
   Of course $(\rho_{S}(t))_{21} = (\rho_{S}(t))_{12}^{*}$ and
   $(\rho_{S}(t))_{22}= 1 -(\rho_{S}(t))_{11}$.
   By diagonalizing the two by two density matrix, we find the
   eigenvalues
   \begin{equation}
   \Lambda_{1}(t)  = \frac{1}{2} + \frac{\sqrt{[(\rho_{S}(t))_{11}
   -(\rho_{S}(t))_{22}]^{2} +
   4(\rho_{S}(t))_{12}(\rho_{S}(t))_{21}}}{2}
   \label{firsteigenvalue}
    \end{equation}
    and
   \begin{equation}
   \Lambda_{2}(t)  = \frac{1}{2} - \frac{\sqrt{[(\rho_{S}(t))_{11}
   -(\rho_{S}(t))_{22}]^{2} +
   4(\rho_{S}(t))_{12}(\rho_{S}(t))_{21}}}{2}.
   \label{secondeigenvalue}
    \end{equation}
 The time evolution of the Tsallis entropy corresponding to the
 mobile entropic index $q$ is given by
 \begin{equation}
 S_{q}(t)  = \frac{1 - \Lambda_{1}(t) ^{q} - \Lambda_{2}(t) ^{q}}{q-1}.
 \label{mobileindex}
 \end{equation}

 The expression of Eq.(\ref{mobileindex}) is not yet suitable for an
 analytical discussion of the problem under study, since it depends on
 integrals definings the terms of Eqs.(\ref{oneone}) and
 (\ref{onetwo}). Those integrals can be easily solved if we make the
 approximation of neglecting $V^{2}$ compared to $E^{2}$. This
 approximation is equivalent to that of disregarding
 the  $\emph{r.h.s.}$ of Eq.(\ref{memoryofstrongrandomness})).
 We thus obtain:
 \begin{equation}
 \rho_{S}(t))_{11} =
 1 - \frac{V}{2(\gamma^{2}-4V^{2})}
 [\gamma(1 - exp(-2Vt/\hbar) - V(1-exp(-2\gamma t/\hbar)]
 \label{simplifiedoneone}
 \end{equation}
 and
 \begin{equation}
 \rho_{S}(t))_{12} =
\frac{i V}{2\gamma}
 [1 - exp(-2\gamma t/\hbar)].
 \label{simplifiedonetwo}
 \end{equation}
With help of Eqs.(\ref{simplifiedoneone}) and (\ref{simplifiedonetwo}) the
time evolution of the Tsallis entropy becomes analytical.
It is interesting to note that if the entropy
$S_{q}(t)$ is plotted for different values of the entropic index
$q$ it results in the same kind of leaf-shape effect
as that given by the numerical results of Section IV (see Fig. \ref{quattro}).
It is possible to prove analytically that the magic value of the
mobile entropic index $q$, $Q$, is $Q = 1/2$.
The numerical leaf-shape effect means that the adoption of the magic
value of $q$ results in a linear increase of entropy as a function of
time after a transient process of time duration $1/\gamma$. Thus, we
make an expansion of $S_{q}(t)$ supplementing the condition
$V>>\gamma$ (which is made necessary to give credibility to the
two-site model) with additional conditions $V t << 1$
and $\gamma t \approx 1$. We thus obtain from Eq.(\ref{mobileindex}),
with Eqs.(\ref{firsteigenvalue}), (\ref{secondeigenvalue}),
(\ref{simplifiedoneone}) and (\ref{simplifiedonetwo}):
\begin{equation}
 S_{q}(t) \approx \frac{q V^{2} t^{2} -(V^{2}t^{2})^{q}}{q -1}.
 \label{timederivative}
 \end{equation}
 It is easy to prove that the linear dependence
 on $t$ is obtained by assigning to $q$ the
 magic value $Q = 1/2$.In fact, in this case we
 derive from Eq.(\ref{timederivative}):
 \begin{equation}
S_{1/2}(t) \approx -\frac{q V^{2} t^{2}}{\hbar^{2}} +
\frac{2Vt}{\hbar} \approx \frac{2Vt}{\hbar}.
\label{nogamma}
 \end{equation}
 This means that the rate of entropy increase is:
  \begin{equation}
\frac{d}{dt} S_{1/2}(t) \approx \frac{2V}{\hbar} .
\label{crucialresult}
 \end{equation}

\begin{figure}
\centerline{
\vbox{
\psfig{figure=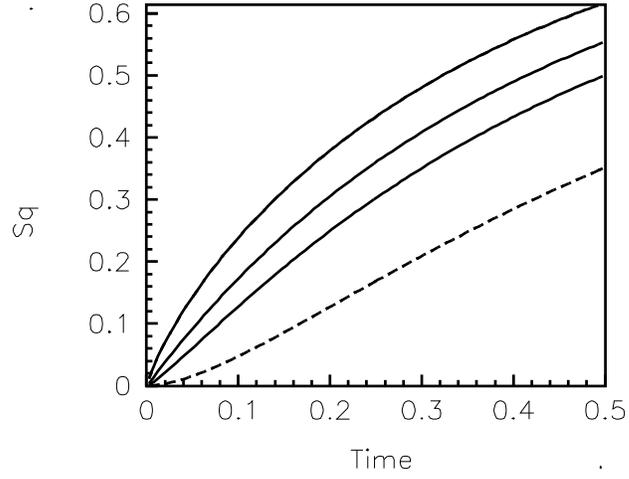,width= 4. in}
}}
\caption[]{
The entropic indicators $S_{q}(t)$ as a function of time with 
Anderson's randomness, $\gamma/\hbar= 100$ and $V/\hbar = 1$.
The plotted curves from the bottom to the top refer to:$q = 1, q = 0.55,
q =0.5, q =0.45$.
}
\label{cinque}
\end{figure}

\begin{figure}
\centerline{
\vbox{
\psfig{figure=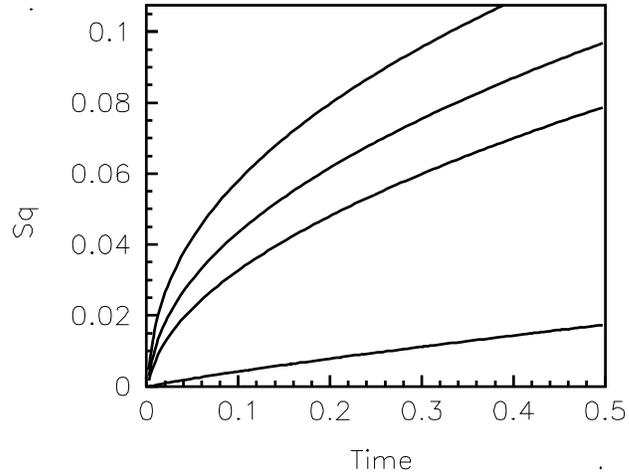,width= 4. in}
}}
\caption[]{
The entropic indicators $S_{q}(t)$ in the presence of only external randomness
as a function of time. The value of system's parameters are $\sigma = 100$ and $V/\hbar = 1$.
The plotted curves from the bottom to the top refer to:$q = 1, q = 0.55,
q =0.5, q =0.45$.
}
\label{sei}
\end{figure}

 It is interesting to notice that the case where the conditions for statistical
 mechanics are
 realized by environmental fluctuations, the two-state model
 discussed in this Section would lead to a master equation identical
 to that of Eq. (\ref{generalizedmasterequation}) with the memory
 kernel of Eq.(\ref{memoryofstrongrandomness}) replaced by:

 \begin{equation}
    K(t) = \frac{4 V^{2}}{\hbar^{2}} \exp(-\frac{\sigma}{\hbar} t).
    \label{externalrandomness}
    \end{equation}
In this specific case
an analytical treatment of the same kind as that illustrated above
yields $Q = 1$.
In Fig. \ref{cinque} and \ref{sei} we compare the entropy time evolution
produced by Anderson's randomness to the externally induced entropy increase.
In fact in Fig.\ref{cinque} we study the leaf effect associated to Anderson's
randomness $\gamma = 100$, changing the mobile
entropic index $q$ from $q = 1$ to $q= 0.45$ .Comparing Fig. \ref{sei} to
Fig.\ref{cinque} we see that the typical leaf effect, induced by Anderson's
randomness, is lost if the entropy increase is only of external origin.
We also note that the leaf effect in this case is
similar to that of Fig.\ref{quattro}, the only remarkable difference being the fact
that the large $\gamma$ condition has the effect of strongly reducing
the time duration of the first time regime.

\begin{figure}
\centerline{
\vbox{
\psfig{figure=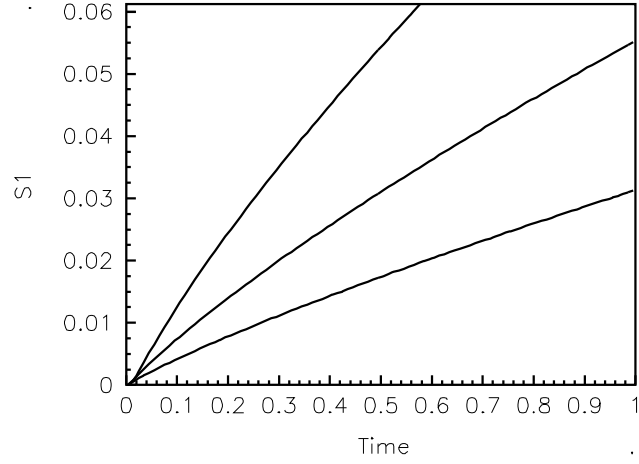,width= 4. in}
}}
\caption[]{
Externally generated time increase of $S_{1}(t)$.
$V/\hbar = 1 $. From the bottom to the top: $\sigma/\hbar= 400,
\sigma/\hbar = 200, \sigma/\hbar = 100$.
}
\label{sette}
\end{figure}

With the help of Figs. \ref{sei} and \ref{sette} we note that in the case of merely external
randomness
the rate of entropy increase is proportional to
$V^{2}/\sigma$. In fact from Figs.\ref{sei} and \ref{sette} we see that the entropy rate is
proportional to $1/ \sigma$ and  to $V^{2}$, respectively.
In other words, we find that in this case the entropy
increase corresponds to the rate of the environment induced
dephasing process. This has to be contrasted with the earlier discovery of
Eq. (\ref{crucialresult}). We see, in other words, that in the case of
merely external randomness the ``magic'' entropic index
is given by the conventional entropic index $Q = 1$ and that the
adoption of this magic entropic index reveals
an ordinary source of randomness. In the case where
the only source of randomness is internal, namely, is
Anderson's randomness, the ``magic'' entropic index is given by
value $Q = 0.5$.The corresponding non-extensive entropy is the proper
entropic indicator signalling that
a thermodynamic view is still possible in spite of
dynamics dominated by strong correlations.

    \section{concluding remarks}

\begin{figure}
\centerline{
\vbox{
\psfig{figure=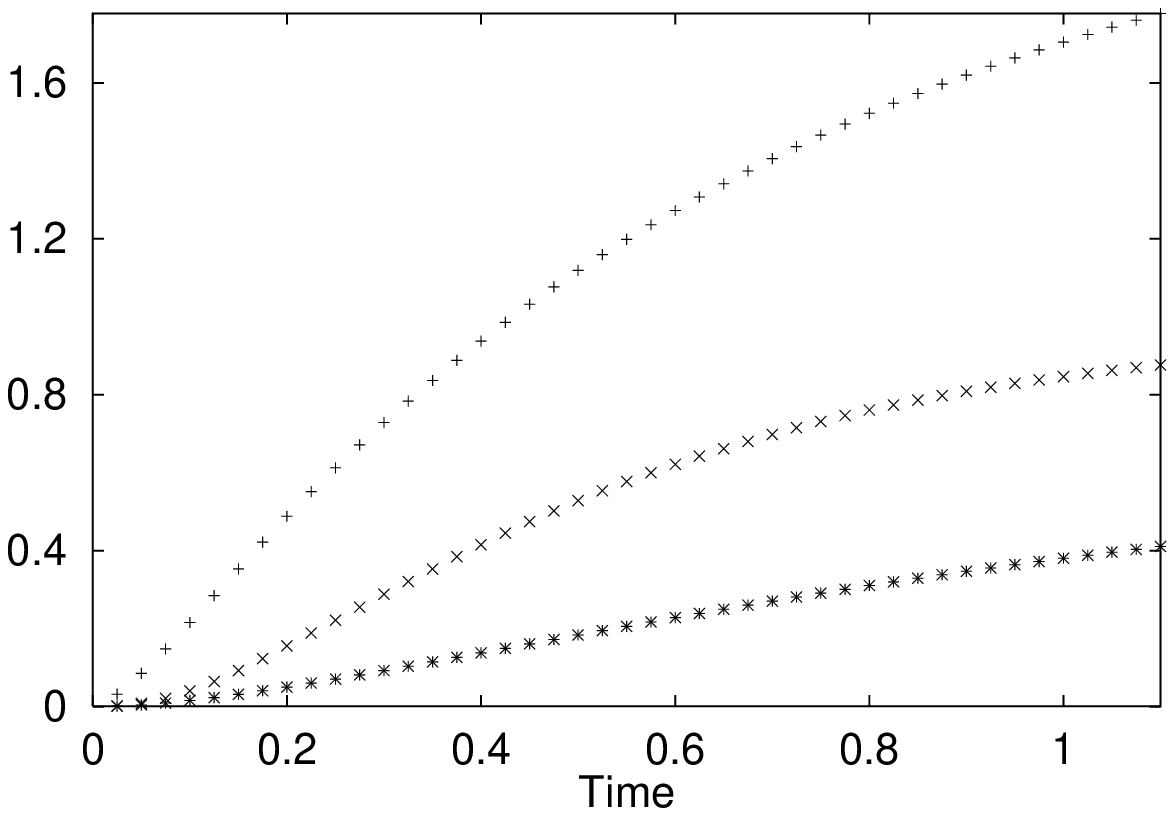,width= 4. in}
}}
\caption[]{
  The entropic indicators and the second moment of diffusion as a
  function of time. $\gamma/\hbar=4$ and $V/\hbar=1$. The curve denoted by
  the $+'s$ is the Tsallis entropy
  with $Q =0.5$; the curve denoted by the $X's$ is the Gibbs entropy (the
  Tsallis entropy with $Q =1$); the curve denoted by the $*'s$ is
  the second moment of the distribution, $M_{2}(t)$. Time is
  expressed in units of $\hbar/V$. 
}
\label{uno}
\end{figure}

\begin{figure}
\centerline{
\vbox{
\psfig{figure=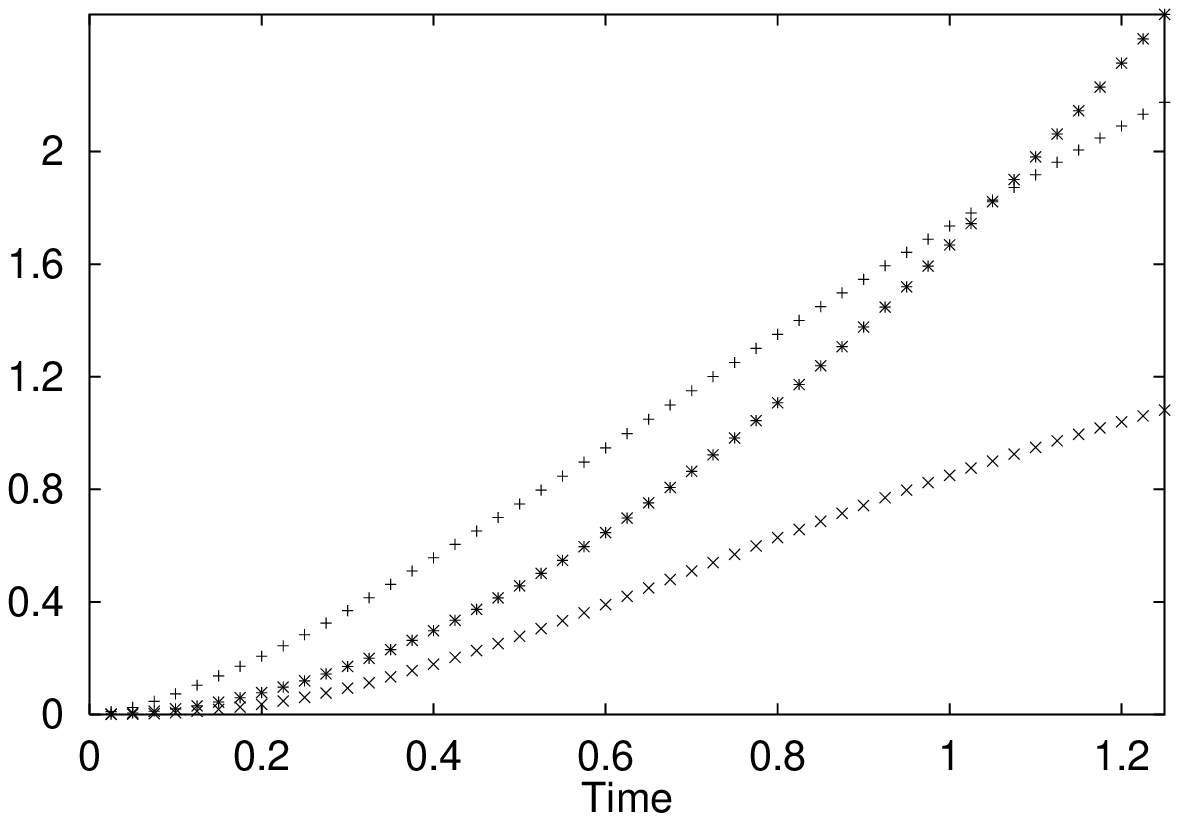,width= 4. in}
}}
\caption[]{
  The entropic indicators and the diffusion second moment as a
  function of time. $\gamma/\hbar=0.3$ and $V/\hbar=1$. The curve denoted by the $+'s$
  is the Tsallis entropy
  with $Q =0.5$; the curve denoted by the $X's$ is the Gibbs entropy (the
  Tsallis entropy with $Q =1$); the curve denoted by the $*'s$ is
  the second moment of the distribution, $M_{2}(t)$. Time is
  expressed in units of $\hbar/V$.

}
\label{due}
\end{figure}

\begin{figure}
\centerline{
\vbox{
\psfig{figure=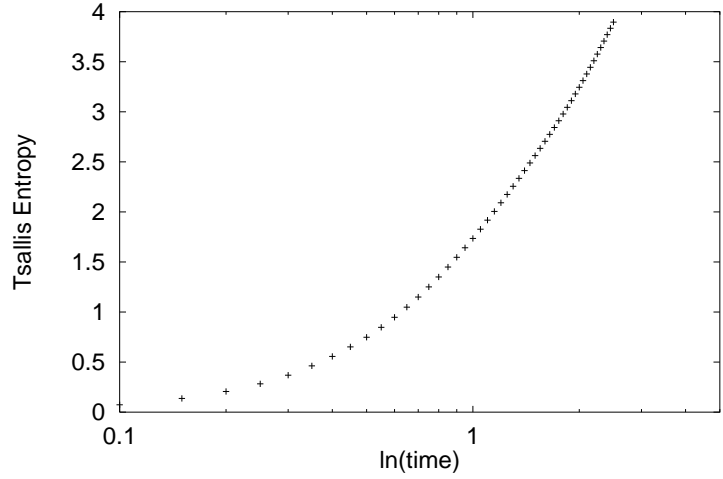,width= 3.8 in}
}}
\caption[]{
  The Tsallis entropy with $q = 0.5$ expressed with respect to $log(t)$
  with the same parameters as those of Fig.\ref{due}
}
\label{tre}
\end{figure}

    The first indisputable result of this paper is given by Figs.\ref{uno},
 \ref{due} and \ref{tre}.
    These figures prove that the the tight-binding Hamiltonian
    of Eq.(\ref{totalhamiltonian}) is a source of entropy time
    evolution clearly showing the three regimes recently discovered by Latora
    and Baranger\cite{baranger}.Furthermore, from Fig.\ref{uno} we see that
    the regime of Brownian diffusion
    discussed by Mazza and Grigolini\cite{mazza}
    is a regime of constant KS entropy.

    On the basis of this observation, we would be tempted to conclude
    that this Kolmogorov regime is characterized by ordinary
    statistical mechanics,  but this interpretation would not be
    not totally satisfactory. In fact, it is well known\cite{mazza}
    that this regime of apparently ordinary statistical mechanics
    is compatible with the silent action of
    quantum correlations. In the  case $\gamma >> V$ this has to do
    with the silent action of the negative tail of
    Eq.(\ref{memoryofstrongrandomness}), whose time scale
    is $\hbar/(2V)$. At the end of this transient process, Anderson's
    localization takes place. Thus, we find to be to some extent
    embarassing to interpret this regime as a manifestation of
     ordinary statistical mechanics.
    According to the illuminating picture illustrated by Zaslavsky in
    his recent book\cite{zaslavsky}, ordinary statistical mechanics are
    closely related to a deterministic motion of the type of the
    Bernouilli shift, map, namely, a case of dynamics with no memory.
    Here, on the contrary, the ensuing process of Anderson's
    localization is a consequence of the action of quantum
    correlations, even if this remained silent through the whole
    Kolmogorov regime of Fig.\ref{uno}.
    For this reason, we find it to be extremely interesting
    that the adoption of the Tsallis entropy makes a different
    interpretation emerge. This is so because the regime
    of steady entropy increase per unit time becomes well distinct only if
    we use the magic value $Q =1/2$, which implies that
    the Brownian diffusion regime studied by Mazza and
    Grigolini\cite{mazza} is actually a form of non-extensive rather
    than extensive thermodynamics. This is compatible with the fact
    that this proces is characterized by the silent action of quantum
    correlations which are responsible for the occurrence of Anderson's
    localization. We note that this interpretation is also supported
    by the observation, numerical and analytical, that the rate of
    entropy increase is $2V/\hbar$. In the case $\gamma >> V$ this is
    also the rate of establishment of Anderson's localization.

    The results illustrated by Fig.\ref{due} are an exciting confirmation of
    the fact that the magic value $Q =0.5$ results in in a marked
    regime of linear entropy increase.However, in this case the onset
    of Anderson localization takes place at a time scale much larger
    than $\hbar/V$, and this makes it difficult to maintain the claim
    about a direct connection between $Q <1$ and Anderson's
    localization. We think that in this specific  case the frequency
    $V/\hbar$ has to do with the coherent motion of a regular lattice.
    The memory kernel $K(t)$ of Eq.(\ref{generalizedmasterequation})
    undergoes many oscillations with frequencies of this order of
    magnitude before producing localization. Thus, in this specific
    case the connection between $Q<1$ and Anderson's localization seems
    to be much less direct.

    Is it possible that $Q<1$ might signal quantum coherent motion,
    without necessarily implying Anderson's localization? We think
    that some more research work must be devoted to this intriguing
    issue. The results of this paper give some more support to the
    conjecture of Ref.\cite{luigi} that $Q <1$ is a signature
    of localization. However, we want to point out that the arguments of
    Section IA cannot yet be used as a compelling evidence that this
    is
    true in general, especially because the arguments of Section IA
     refer to the classical
    case of fractal dynamics.
    We are convinced that there exists a statistical equivalence
    betwen fractal dynamics and Anderson's randomness. We hope that
    the results of this paper might trigger further research to prove,
    or disprove, this interesting conjecture.

     Although this conjecture is not yet proved, it cannot be easily
     dismissed either.
    From the theory of Section IV
    we see that if Anderson's randomness
    vanishes, and entropy increase only rests on external fluctuations,
    the magic entropic index $Q$ is given the ordinary value $Q=1$,
    again. On top of that, the rate of entropy increase is
    $V^{2}/\gamma$, in accordance with the fact that entropy measures
    randomness, and that the rate of decoherence
    is a proper measure of system's randomness. Thus, we can conclude
    that the adoption of Tsallis entropy makes it possible to adopt
    a thermodynamics perspective, even when the entropy rate signals a
    coherent property, $2V/\hbar$, rather than an incoherent dephasing
    process, and that the miracle is possible due to the adoption of
    the entropic index $Q = 0.5$.

\newpage

\begin{figure}
\centerline{
\vbox{
\psfig{figure=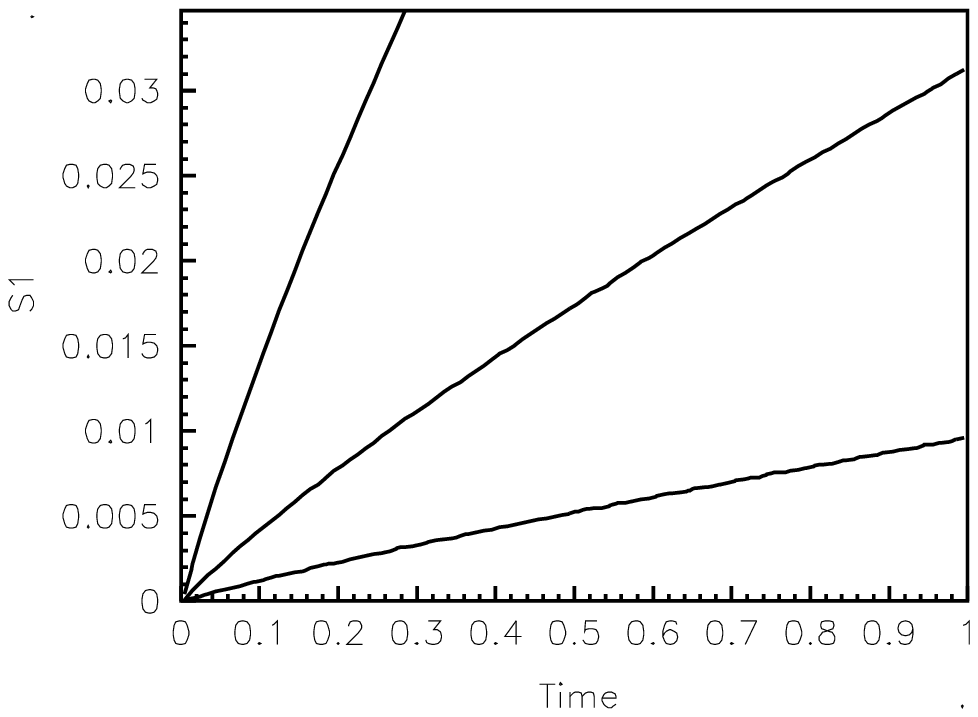,width= 4. in}
}}
\caption[]{
Externally generated time increase of $S_{1}(t)$.
 $\sigma/\hbar = 400$. From the bottom to the top:  $V/\hbar= 0.5,
V/\hbar = 1, V/\hbar = 2$.
}
\label{otto}
\end{figure}


\begin{thebibliography}{99}
\bibitem{note1}
 Actually, in the real dynamic cases, the sensivity to initial condition
 does not result in a
 simple expression as that of Eq. (\ref{typicalsensitivity}). Rather,
 as earlier found by the authors of Ref.\cite{politi}, in the specific case of the logistic map
 a distribution of power indexes $\beta = \frac{1}{1-Q}$. There seem to be good reasons to believe
 the proper entropic index fit the condition $Q<1$.
 is found. However, as pointed out in Ref.\cite{simone}, this
 distribution ranges from the value
 \begin{equation}
 \frac {1}{1-Q} = \frac{1}{\alpha_{min}}- \frac{1}{\alpha_{max}}
 \label{tsallisprediction}
 \end{equation}
 to the value $\beta = 0$ thereby always implying the condition $Q <1$
 to be fulfilled.
\end{thebibliography}
\end{document}